\documentclass[aps,showpacs,twocolumn,pra,superscriptaddress]{revtex4}

\UseRawInputEncoding

\usepackage[tbtags]{amsmath}
\usepackage{amssymb}
\usepackage{amsfonts}
\usepackage{bmpsize}
\usepackage{mathtools}
\usepackage{amsthm}
\usepackage{mathrsfs}
\usepackage[colorlinks=true,citecolor=blue,urlcolor=blue]{hyperref}
\hypersetup{colorlinks=true,linkcolor=blue,filecolor=blue,urlcolor=blue,citecolor=blue}
\usepackage{multirow}

\begin{document}
\title{Network nonlocality sharing via weak measurements in the extended bilocal scenario}

\author{Wenlin Hou}
\affiliation{Key Laboratory of Low-Dimensional Quantum Structures and Quantum Control of Ministry of Education, Key Laboratory for Matter Microstructure and Function of Hunan Province, Department of Physics and Synergetic Innovation Center for Quantum Effects and Applications, Hunan Normal University, Changsha 410081, China}

\author{Xiaowei Liu}
\affiliation{Key Laboratory of Low-Dimensional Quantum Structures and Quantum Control of Ministry of Education, Key Laboratory for Matter Microstructure and Function of Hunan Province, Department of Physics and Synergetic Innovation Center for Quantum Effects and Applications, Hunan Normal University, Changsha 410081, China}

\author{Changliang Ren}\thanks{Corresponding author: renchangliang@hunnu.edu.cn}
\affiliation{Key Laboratory of Low-Dimensional Quantum Structures and Quantum Control of Ministry of Education, Key Laboratory for Matter Microstructure and Function of Hunan Province, Department of Physics and Synergetic Innovation Center for Quantum Effects and Applications, Hunan Normal University, Changsha 410081, China}

\begin{abstract}
Quantum network correlations have attracted strong interest as the emergency of the new types of violations of locality. Here we investigated network nonlocal sharing in the extended bilocal scenario via weak measurements. Interestingly, network nonlocal sharing can be revealed from the multiple violation of BRGP inequalities of any $\mathrm{Alice}_n-\mathrm{Bob}-\mathrm{Charlie}_m$, which has no counterpart in the case of Bell nonlocal sharing scenario. Such discrepancy may imply an intrinsic difference between network nonlocality and Bell nonlocality. Passive and active network nonlocal sharing as two distinct types of nonlocality sharing have been completely analyzed. In addition, noise resistance of network nonlocal sharing were also discussed.
\end{abstract}


\maketitle

\section{Introduction}

The non-classical characteristic of quantum physics was firstly concerned by Einstein et.al\cite{Einstein} in their pioneering paper, which shows that there are some conflicts between quantum mechanics and local realism. With the continuous in-depth research, physicists realized that understanding such incompatibility needs to answer whether the local hidden variable theory (LHV) can explain the predictions of quantum mechanics or not. However, there were no empirical ideas or paths to solve this problem until 1964, when Bell showed that in all local realistic theories, correlations between the outcomes of measurements in different parts of a physical system satisfy a certain class of inequalities \cite{Bell}. In contrast, it is easy to find that entangled states violate these inequalities \cite{ Bell, Clauser, Zukowski1, Mermin, Belinskii, Ardehali, Collins, Brukner} in quantum mechanics, which clearly shows the crucial conflict between classical theory and quantum mechanics. Hence, Bell's work was described as ``the most profound discovery of science" \cite{Stapp}. Aspect \emph{et al.} experimentally observed the violation of Bell's inequality in 1981 \cite{Aspect} for the first time. Following the deep exploration of nonlocality, it leads to the birth of quantum information. Nonlocality is crucial for our understanding of quantum mechanics and represents a resource in device-independent quantum information protocols, including quantum key distribution \cite{Acin}, quantum computation \cite{Horodecki,Brunner}, quantum metrology \cite{Giovannetti}, and random number generation \cite{Pironio, Colbeck}, etc. All these studies mainly involve the same ``Bell scenario", that is, Bell nonlocality \cite{Brunner}.

However, in the past decade, physicists began to explore more complex quantum network \cite{Kimble, Wehner}. Whether there exist new kinds of quantum nonlocality different from Bell nonlocality is a fundamental topic. A type of network nonlocality which may beyond Bell nonlocality has been gradually revealed \cite{Branciard,Fritz,Branciard-1,Henson,Renou,Bancal,berg,Tavakoli-1,Coiteux,Tavakoli}. The key idea of network nonlocality is that, different sources that distribute their physical system to the nodes of the network should be independent of each other, which represents a fundamental departure from the standard Bell nonlocality \cite{Branciard}. At present, the investigations on network nonlocality mostly focus on the bilocal scenario and the triangle network scenario. In the "bilocal" scenario, two independent sources share entangled pairs with three observers, which is of particular relevance since it corresponds to the scenario underlying entanglement swapping experiments \cite{Zukowski2}. And the triangle network scenario involves a triangular geometry: three independent sources on the edge of a triangle distribute three entangled states to three measuring devices on the corner of the triangle. In these years, some progress has been made in the research of network nonlocality. For instance, Renou et al. proved that quantum network with a triangular geometry displays nonclassical correlations that appear to be fundamentally different from those so far revealed through Bell tests \cite{Renou}. Nevertheless, the investigation of network nonlocality is still in it's infancy. This rapidly developing topic presently finds itself at a point where several basic methods and tools for its systematic analysis are being established, and many elementary questions remain wide open. So far, our understanding of quantum correlation in networks is still very limited. One of the most important difficulties is how to exhibit that the network nonlocality is fundamentally different from Bell nonlocality.

In 2015, a surprising result was reported\cite{Silva}, showing that Bell nonlocality of a pair of qubits may be actually shared among more than two observers using weak measurements, and experimentally demonstrated by refs. \cite{Hu, Schiavon, Feng} later. Nonlocal sharing has attracted extensive attention \cite{Ren,DAS,Sasmal,Bera,Datta,Shenoy,Kumari,Saha,Mohan,Roy,Srivastava,Kanjilal,Yao,Zhu,Cheng,ren-1}, and lots of relevant results have been published ever since, such as active and passive nonlocal sharing\cite{Ren}, quantum steering sharing\cite{Yao} and so on. However, almost all discussions are limited to the case of one-sided sequential measurements, that is, an entangled pair is distributed to one Alice and multiple Bobs. Recently, Zhu et al. explored the nonlocal sharing in bilateral sequential measurements\cite{Zhu,Cheng,ren-1}, in which a pair of entangled states is distributed to multiple Alices and Bobs. It is shown that nonlocality sharing between $\mathrm{Alice_1}$-$\mathrm{Bob_{1}}$ and $\mathrm{Alice_2}$-$\mathrm{Bob_{2}}$ is impossible in this case. At present, all nonlocal sharing discussions focus on Bell-type quantum nonlocality, and there is no investigation on network nonlocality. It is still unclear whether network non-locality can be shared with the help of weak measurements or not. In particular, can it exhibit the fundamental difference between Bell nonlocality and network nonlocality, and provide more evidence?

Inspired by the above problems, we investigated network nonlocal sharing based on weak measurement in the extended bilocal scenario. Interestingly, network nonlocal sharing can be revealed from the multiple violation of BRGP inequalities of any $\mathrm{Alice}_n-\mathrm{Bob}-\mathrm{Charlie}_m$, which has no counterpart in the case of Bell nonlocal sharing scenario. Because it is impossible to observe either the nonlocality sharing between $\mathrm{Alice_1}$-$\mathrm{Bob_{1}}$ and $\mathrm{Alice_2}$-$\mathrm{Bob_{2}}$ or the nonlocality sharing between $\mathrm{Alice_1}$-$\mathrm{Bob_{2}}$ and $\mathrm{Alice_2}$-$\mathrm{Bob_{1}}$ in Bell nonlocal sharing scenario \cite{Zhu,Cheng}. Such discrepancy may give a hint to exhibit the different properties between network nonlocality and Bell nonlocality. Similarly, network nonlocal sharing can also be divided into two different types of nonlocality sharing, passive and active network nonlocal sharing, according to the different motivations of the former observers $\mathrm{Alice_1}$ and $\mathrm{Charlie_1}$. Active network nonlocal sharing
can be observed even when $\mathrm{Alice_1}$ and $\mathrm{Charlie_1}$ perform strong measurement except the ideal strong measurement. In addition, the noise resistance of such network nonlocal sharing has also been discussed. These results not only may shed light on the difference between network nonlocality and Bell nonlocality, but also can be applied in some quantum network tasks.

The structure of the paper is as follows. In Sec. \ref{framework}, we primarily introduced the framework of extended bilocal scenario and operation process. Passive network nonlocal sharing and active network nonlocal sharing are completely analyzed in Sec. \ref{passive} and Sec. \ref{active}. Noise resistance of such network nonlocal sharing were discussed in Sec. \ref{noise}.

\begin{figure}[htbp]\label{Scenario}
      \centering
      \includegraphics[width=0.4\textwidth]{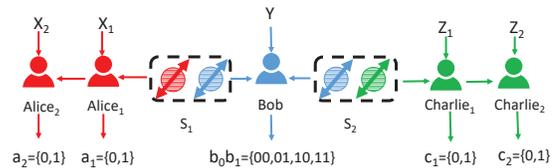}
      \caption{\small{The extended bilocal scenario: a typical entanglement swapping scenario includes $\mathrm{Alice_1}$, Bob and $\mathrm{Charlie_{1}}$. A two-qubit entangled state produced by the source $S_{1}$ is shared by $\mathrm{Alice_1}$ and Bob and  another two-qubit entangled state produced by the source $S_{2}$ is shared by Bob and $\mathrm{Charlie_1}$. Unlike this, the extended bilocal scenario has another two observers $\mathrm{Alice_2}$ and $\mathrm{Charlie_2}$. When $\mathrm{Alice_1}$ has measured her particle she pass it to $\mathrm{Alice_2}$. $\mathrm{Charlie_2}$ receives the particle which has been weak-measured by $\mathrm{Charlie_1}$ similarly.
      }} \label{Bilocal Scenario}
    \end{figure}

\section{The framework of the extended bilocal scenario }\label{framework}

As illustrated by Fig.1, the scenario that we discussed is extended from the standard bilocal scenario. There are two sources where each of them generates a 2-qubit state. One source sends particles to Alices and Bob, and another source sends particles to Charlies and Bob. Different from the standard bilocal scenario, two observers, $\mathrm{Alice}_1$ and $\mathrm{Alice}_2$, will measure their shared particle sequentially, similarly for $\mathrm{Charlie}_1$ and $\mathrm{Charlie}_2$ in the other side. In particular, each observer $\mathrm{Alice}_n$ ($\mathrm{Charlie}_m$) chooses two different dichotomic observables independently, denoted by $\hat{X}_n\in\{\hat{A}_{n,i}\}$ ($\hat{Z}_m\in\{\hat{C}_{m,j}\}$) with binary outcome $a_n\in\{a_{n,i}\}$($c_m\in\{c_{m,j}\}$), where $n,m\in\{1,2\}$, and $i,j,a_{n,i},c_{m,j}\in\{0,1\}$. Bob performs a complete
bell state measurement defined as $\hat{Y}$ with four distinguishable outputs $b\equiv b_{0}b_{1}\in\{00,01,10,11\}$. The joint probability distribution of the measurement outcomes of these observables can be described as $P(a_1,a_2,b,c_1,c_2|\hat{X}_1,\hat{X}_2,\hat{Y},\hat{Z}_1,\hat{Z}_2)$.

According to the complete joint probability distribution, we can always obtain the joint probability of any combination of $\mathrm{Alice}_n-\mathrm{Bob}-\mathrm{Charlie}_m$, which can be given as,
\begin{eqnarray}\label{Bilocal}
&&P(a_{n},b,c_{m}\mid \hat{X}_{n},\hat{Y},\hat{Z}_{m})=
\frac{1}{4} \nonumber\\
&&\sum_{\mbox{\tiny $\begin{subarray}{l}
a_{n'},c_{m'},\hat{X}_{n'},\hat{Z}_{m'}\\
n\neq n',m\neq m'\end{subarray}$}}P(a_{1},a_{2},b,c_{1},c_{2}|\hat{X}_1,\hat{X}_2,\hat{Y},\hat{Z}_1,\hat{Z}_2)
\end{eqnarray}
Hence, it is possible to analyze quantum nonbilocal correlations among of three different observers, $\mathrm{Alice_n}$, Bob and $\mathrm{Charlie_{m}}$. Based on the assumption, the tripartite distribution
$P(a_{n},b,c_{m}\mid \hat{X}_{n},\hat{Y},\hat{Z}_{m})$ is local when it can be written in the factorized form

\begin{eqnarray}\label{Bell state}
&&P(a_{n},b,c_{m}| \hat{X}_{n},\hat{Y},\hat{Z}_{m})=\int_{\lambda_1}\mathrm{d}\lambda_1 \int_{\lambda_2} \mathrm{d}\lambda_2 \ p(\lambda_1) p(\lambda_2) \nonumber \\
&& \times p(a_{n} |\hat{X}_{n}, \lambda_1) p(b | \hat{Y}, \lambda_1, \lambda_2) p(c_{m} | \hat{Z}_{m}, \lambda_1)
\end{eqnarray}
where the two sets of distributions of hidden variables $\lambda_1$ and $\lambda_2$ are originated from two independent sources. Clearly, the local response function for $\mathrm{Alice_n}$ only depends on
$\lambda_1$ and the one of $\mathrm{Charlie_{m}}$ only depends on $\lambda_2$, while the one of Bob depends on both $\lambda_1$ and $\lambda_2$. In particular, if $\mathrm{Alice_1}$ and $\mathrm{Charlie_{1}}$ perform strong measurements, either $\mathrm{Alice_2}$ or $\mathrm{Charlie_{2}}$ will receive a projective eigenstate, where any correlation in the origin source has been destroyed completely. Since $\mathrm{Alice_1}$ and $\mathrm{Alice_2}$ ($\mathrm{Charlie_{1}}$ and $\mathrm{Charlie_{2}}$) choose measurements independently without communication, the tripartite distribution
$P(a_{n},b,c_{m}\mid \hat{X}_{n},\hat{Y},\hat{Z}_{m})$ for any $\mathrm{Alice}_2-\mathrm{Bob}-\mathrm{Charlie}_m$ or $\mathrm{Alice}_n-\mathrm{Bob}-\mathrm{Charlie}_2$ is local, and can be always decomposed in the form of Eq.(\ref{Bilocal}). Hence, it is trivial to further consider the later observers $\mathrm{Alice}_2$ and $\mathrm{Charlie}_2$, while returns to the standard bilocal scenario. In this extended scenario, $\mathrm{Alice}_1$ and $\mathrm{Charlie}_1$ will perform weak measurements, whereas Bob, $\mathrm{Alice}_2$ and $\mathrm{Charlie}_2$ will carry out strong measurements.

Without loss of generality, it is necessary to obtain the explicit results of the tripartite distribution $P(a_{n},b,c_{m}\mid \hat{X}_{n},\hat{Y},\hat{Z}_{m})$ for any $\mathrm{Alice}_n-\mathrm{Bob}-\mathrm{Charlie}_m$ in the extended bilocal scenario. Supposed that the two shared states emitted by the two sources $S_1$ and $S2$ are defined as $\rho_{\mathrm{AB}}$ and $\rho_{\mathrm{BC}}$ respectively, the whole initial state of the system can be described by,
\begin{eqnarray}\label{network state1}
\rho_{ABC}=\rho_{AB}\otimes \rho_{BC}.
\end{eqnarray}
Bob carries out a Bell state measurement on the two particles he receives, with four possible outputs $b=b_0b_1=\{00,01,10,11\}$ corresponding to the four Bell states $\{|\phi^+\rangle, |\phi^-\rangle,|\psi^+\rangle,|\psi^-\rangle\}$ respectively, where $|\phi^\pm\rangle=\frac{1}{\sqrt{2}}(|00\rangle\pm|11\rangle)$ and $|\psi^\pm\rangle=\frac{1}{\sqrt{2}}(|01\rangle\pm|10\rangle)$. As a matter of convenience, we defined the density matrices of these four Bell state as $\rho_{b_{0}b_{1}}$ which we will used in later discussion. The observers, $\mathrm{Alice_1}$ and $\mathrm{Alice_2}$, will measure their shared particle sequentially, similarly for $\mathrm{Charlie_1}$ and $\mathrm{Charlie_2}$ in the other side. In the scenario, the first observer of each side performs the optimal weak measurements, while the second observer of each side carries out strong measurements. Without of loss of generality, each of the observers chooses two different dichotomic operators independently, which can be defined as
\begin{eqnarray}\label{Alice M}
\hat{A}_{n,i}=\cos\theta_{n,i}\sigma_z-(-1)^{i}\sin\theta_{n,i}\sigma_{x}
\end{eqnarray}
and
\begin{eqnarray}\label{Charlie M}
\hat{C}_{m,j}=\cos\theta_{m,j}\sigma_{z}+(-1)^{j}\sin\theta_{m,j}\sigma_{x}
\end{eqnarray}
where $\sigma_{x}$ and $\sigma_{z}$ are pauli matrices.

When Bob performs a Bell state measurement with the result $b_{0}b_{1}$ on two particles he has received, the state of the whole system will change to
\begin{eqnarray}\label{Bell state8}
\rho^{b_{0}b_{1}}_{\mathrm{ABC}}=(\mathbb{I}\otimes\rho_{b_{0}b_{1}}\otimes \mathbb{I}). \rho_{\mathrm{ABC}}.(\mathbb{I}\otimes\rho_{b_{0}b_{1}}\otimes \mathbb{I})^ \dagger.
\end{eqnarray}
Thereby, after Bob's measurement, the reduced state on Alice and Charlie's side can be obtained by tracing over Bob's system,
\begin{eqnarray}\label{Bell state9}
\rho^{b0b1}_{AC}=tr_{B}(\rho^{b0b1}_{ABC}).
\end{eqnarray}
Here $\rho^{b_{0}b_{1}}_{AC}$ is not normalized. Then, $\mathrm{Alice_1}$ performs a weak measurement
on her subsystem, the reduced state can be given as,
\begin{align}
\rho_{\hat{X}_1}^{a_1}=&\frac{F_1}{2}\rho^{b0b1}_{AC}+\frac{1+(-1)^{a_1}G_1-F_1}{2}[U_{\hat{X}_1}^{1}\rho^{b0b1}_{AC}(U_{\hat{X}_1}^{1})^ \dagger]\nonumber \\ &+\frac{1-(-1)^{a_1}G_1-F_1}{2}[U_{\hat{X}_1}^{0} \rho^{b0b1}_{AC}(U_{\hat{X}_1}^{0})^ \dagger]
\end{align}
where $U_{\hat{X}_n}^{a_{n}}=\Pi_{\hat{X}_n}^{a_{n}}\otimes \mathbb{I}$ and $\Pi_{\hat{X}_n}^{a_{n}}\in\{\frac{\mathbb{I}+(-1)^{a_{n,i}} A_{n,i}}{2}\}$. $F_{1}$ is the quality factor which represents the undisturbed
extent to the state of $\mathrm{Alice_1}$'s qubit after she measured and $G_1$ is the precision factor which quantifies the information gain from $\mathrm{Alice_1}$'s measurements. Subsequently, $\mathrm{Alice_2}$ performs a strong measurement $\hat{X}_2$ with the outcome $a_2$, the state changed to\begin{align}
\rho_{\hat{X}_2}^{a_2}=U_{\hat{X}_2}^{a_2}\rho_{\hat{X}_1}^{a_1} ({U_{\hat{X}_2}^{a_2}})^ \dagger.
\end{align}

Similarly, $\mathrm{Charlie_1}$ performs weak measurements $\hat{Z}_{1}$ on his received qubit with the quality factor $F_2$ and the precision factor $G_2$ of the measurements. When
the measurement outcome is $c_1$, then the reduced state can be described as
\begin{align}
\rho_{\hat{Z}_1}^{c_1}=&\frac{F_2}{2}\rho_{\hat{X}_2}^{a_2}+\frac{1+(-1)^{c_1}G_2-F_2}{2}[U_{\hat{Z}_1}^{1}\rho_{\hat{X}_2}^{a_2}(U_{\hat{Z}_1}^{1})^ \dagger]\nonumber \\ &+\frac{1-(-1)^{c_1}G_2-F_2}{2}[U_{\hat{Z}_1}^{0} \rho_{\hat{X}_2}^{a_2}(U_{\hat{Z}_1}^{0})^ \dagger]
\end{align}
where $ U_{\hat{Z}_m}^{c_{m}}=\mathbb{I} \otimes\Pi_{\hat{Z}_m}^{c_{m}}$ and $\Pi_{\hat{Z}_m}^{c_{m}}\in\{\frac{\mathbb{I}+(-1)^{c_{m,j}} C_{m,j}}{2}\}$. Finally, $\mathrm{Charlie_2}$ performs a strong measurement $\hat{Z}_2$ with the outcome $c_2$, the reduced state will change to
\begin{align}
\rho_{\hat{Z}_2}^{c_2}=U_{\hat{Z}_2}^{c_2}\rho_{\hat{Z}_1}^{c_1} ({U_{\hat{Z}_2}^{c_2}})^ \dagger.
\end{align}
Therefore, according to the unnormalized postmeasurement state $\rho_{\hat{Z}_2}^{c_2}$, we can obtain the joint conditional probability distribution, which is
$P(a_{1},a_{2},b_{0},b_{1},c_{1},c_{2}\mid \hat{X}_{1},\hat{X}_{2},\hat{Z}_{1},\hat{Z}_{2})=\mathrm{Tr}[\rho_{\hat{Z}_2}^{c_2}]$. In the whole measurement process, the measurements of all observers are completely independent, and  the measurement choices are completely unbiased both for $\mathrm{Alice_n}$ and $\mathrm{Charlie_m}$. By summing up the irrelevant parties, the tripartite distribution
$P(a_{n},b,c_{m}\mid \hat{X}_{n},\hat{Y},\hat{Z}_{m})$ for any $\mathrm{Alice}_2-\mathrm{Bob}-\mathrm{Charlie}_m$ or $\mathrm{Alice}_n-\mathrm{Bob}-\mathrm{Charlie}_2$ can be obtained by Eq.(\ref{Bilocal}).

Here it is reasonable to emphasize that only the two observers on the same side should measure sequentially, such as $\mathrm{Alice_1}$ should measured before $\mathrm{Alice_2}$, the same for $\mathrm{Charlie}_1-\mathrm{Charlie}_2$. However, there is no assumption about the measurement order of different-side observers. In other words, the sequence of local measurements between the observers in different sides does not change the final joint conditional probability distribution $P(a_{1},a_{2},b_{0},b_{1},c_{1},c_{2}\mid \hat{X}_{1},\hat{X}_{2},\hat{Z}_{1},\hat{Z}_{2})$. For simplicity, we follow the sequence of $\mathrm{Bob}-\mathrm{Alice}_1-\mathrm{Alice}_2-\mathrm{Charlie}_1-\mathrm{Charlie}_2$ to introduce the measurement process.

To characterize quantum nonbilocal correlations among three different observables, $\mathrm{Alice_n}$, Bob and $\mathrm{Charlie_{m}}$, it is reasonable to check whether the nonlinear bilocal inequality (BRGP) can be violated or not, which is derived from the bilocality assumption. Once BRGP inequality is violated, it will exclude any possible bilocal model for a tripartite distribution. The BRGP inequality for any combination of $\mathrm{Alice}_n-\mathrm{Bob}-\mathrm{Charlie}_m$ can be given as,
\begin{eqnarray}\label{BRGP}
\sqrt{\mid I_{n,m} \mid}+\sqrt{\mid J_{n,m} \mid} \leq 1
\end{eqnarray}
where
\begin{eqnarray}
I_{n,m}=\frac{1}{4}\sum_{i,j=0,1}\langle \hat{X}_{n}\hat{Y}_{0}\hat{Z}_{m}\rangle
\end{eqnarray}
and
\begin{eqnarray}
J_{n,m}=\frac{1}{4}\sum_{i,j=0,1}(-1)^{i+j}\langle \hat{X}_{n}\hat{Y}_{1}\hat{Z}_{m}\rangle.
\end{eqnarray}
Each tripartite correlation terms can be obtained from the joint probability distribution,
\begin{eqnarray}
\langle \hat{X}_{n}\hat{Y}_{k}\hat{Z}_{m}\rangle=\sum_{\tiny{a_n,b,c_m}}(-1)^{a_n+b_k+c_m}
P(a_n,b,c_m| \hat{X}_{n},\hat{Y},\hat{Z}_{m})\nonumber
\end{eqnarray}
We defined the quantity of the left side of Eq.(\ref{BRGP}), $B_{nm}=\sqrt{\mid I_{n,m} \mid}+\sqrt{\mid J_{n,m} \mid}$. Obviously, in this scenario, there are four BRGP inequalities which can be discussed for different $n$ and $m$.

\section{passive network nonlocal sharing In the extended bilocal scenario)}\label{passive}

Without loss of generality, assumed that the sources $S_1$ and $S_2$ send pairs of particles in the maximal entangled state, $\rho_{\mathrm{AB}}=\rho_{\mathrm{BC}}=|\psi\rangle\langle\psi|$, $|\psi\rangle=\frac{1}{\sqrt{2}}(|00\rangle+|11\rangle)$. Depending on Bob's results, Alice and Charles's particles end up in the corresponding Bell state. As mentioned above, we aim to check whether network nonlocality sharing  between the two sides exists or not in this scenario. Interestingly, similar to the case of Bell's nonlocal sharing scenario, a network  nonlocal sharing investigation can be divided into two types according to the different motivations of the observers $\mathrm{Alice_1}$ and $\mathrm{Charlie_1}$, passive and active network nonlocal sharing. Because, no matter which side, the different measurement choices of the former observer will affect the measurement results of the later observer. Obviously, every observer will choose different independent measurement choices based on their different motivations. Firstly, we discuss the passive network nonlocal sharing. In this case, $\mathrm{Alice_1}$ and $\mathrm{Charlie_1}$ have no conscious thought of nonlocality sharing with the later observers, but only want to achieve a maximal BRGP violation of themselves. The maximal BRGP quantity $B_{11}$ is only limited by the precision factor $G_{1}$ and $G_{2}$, compared to that when $\mathrm{Alice_1}$ and $\mathrm{Charlie_1}$ perform strong measurements.

\begin{figure}[htbp]
      \centering
      \includegraphics[width=0.45\textwidth]{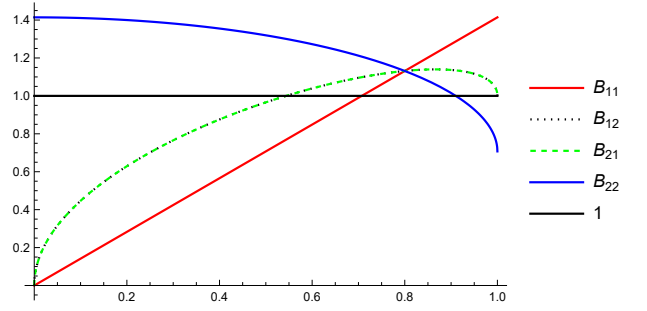}
      \caption{\small{Plot of BRGP quantities $B_{nm} \in \{B_{11}$,$B_{22}$,$B_{12}$,$B_{21}\}$ with red line, blue line, green dashed line and black dotted line respectively for $\mathrm{Alice_{n}-Bob-Charlie_{m}}$ as functions of the precision factor G under the condition of $G_1$ = $G_2$ =$G$. $B_{nm}$ all exceed the bound of 1 in a narrow range.}\label{BAll}}
\end{figure}

$\mathrm{Alice_1}$ and $\mathrm{Charlie_1}$ will select the optimal measurements to achieve the maximal value of $B_{11}$. Using the optimal solution, we can obtain the maximal four BRGP quantities, which are
\begin{eqnarray}\label{B1}
&&B_{11}=\sqrt{2G_{1}G_{2}}\qquad\quad \nonumber
B_{12}=\sqrt{1+F_{2})G_{1}}\nonumber\\
&&B_{21}=\sqrt{(1+F_{1})G_{2}}\quad \nonumber
B_{22}=\sqrt{\frac{(1+F_{1})(1+F_{2})}{2}}
\end{eqnarray}
where the optimal settings are $\theta_{1,0}=\theta_{1,1}=\frac{\pi}{4}$ for $\mathrm{Alice_1}$ and $\mathrm{Charlie_1}$, and $\theta_{2,0}=\theta_{2,1}=\frac{\pi}{4}$ for $\mathrm{Alice_2}$ and $\mathrm{Charlie_2}$. Obviously, when these four BRGP quantities exceed 1 simultaneously, the passive network nonlocal sharing will be observed. For the weak measurements, there are two typical pointer distributions, the optimal pointer and the
square pointer, where the relation between the quality factor $F$ and the precision factor $G$ satisfies $G^2+F^2=1$ or $G+F=1$ respectively with $G,F\in [0,1]$. Hence, the relation between the quality factor $F_1$ and the precision factor $G_1$ can be given based on the pointer type that $\mathrm{Alice_1}$ chooses, and it's similar for $F_2$ and $G_2$ of $\mathrm{Charlie_1}$'s measurements .

As illustrated in Fig.2, $B_{11}=\sqrt{2}G$, $B_{12}=B_{21}=\sqrt{(1+F)G}$ and $B_{22}=\frac{1+F}{\sqrt{2}}$, when the pointer distribution of the weak measurements for $\mathrm{Alice_1}$ and $\mathrm{Charlie_1}$ is optimal and $G_{1}=G_{2}=G$. It is clearly shown that all the four BRGP quantities can exceed 1 simultaneously in the range of $G\in (\frac{1}{\sqrt{2}},\sqrt{2(\sqrt{2}-1)})$, which means network nonlocality can be revealed from the measurement results of any $\mathrm{Alice}_n-\mathrm{Bob}-\mathrm{Charlie}_m$. Such properties of nonlocal sharing are impossible in the Bell nonlocal sharing scenario, where either the nonlocality sharing between $\mathrm{Alice_1}$-$\mathrm{Bob_{1}}$ and $\mathrm{Alice_2}$-$\mathrm{Bob_{2}}$ or the nonlocality sharing between $\mathrm{Alice_1}$-$\mathrm{Bob_{1}}$ and $\mathrm{Alice_2}$-$\mathrm{Bob_{2}}$ never happen. The maximal network nonlocality sharing exists when $G=0.8$ with the maximal violation $1.13137$ for all four BRGP inequalities.

When the pointer distribution of the weak measurements for $\mathrm{Alice_1}$ and $\mathrm{Charlie_1}$ is square, the network nonlocal sharing between $\mathrm{Alice}_1-\mathrm{Bob}-\mathrm{Charlie}_1$ and $\mathrm{Alice}_2-\mathrm{Bob}-\mathrm{Charlie}_2$ can not be observed. As illustrated in Fig.3, $B_{11}^s$ and $B_{22}^s$ can not exceed 1 simultaneously, where the maximum value that these two quantities can achieve at the same time is $0.9428$ when $G_{1}=G_{2}=\frac{2}{3}$. However, when one of the pointer distributions of the weak measurements for $\mathrm{Alice_1}$ and $\mathrm{Charlie_1}$ is optimal and another is square, a double violation of BRGP inequalities $B_{11}^m$ and $B_{22}^m$ between $\mathrm{Alice}_1-\mathrm{Bob}-\mathrm{Charlie}_1$ and $\mathrm{Alice}_2-\mathrm{Bob}-\mathrm{Charlie}_2$ exists. The maximum value that these two quantities can achieve at the same time is 1.034 when $G_{1}=0.702$ and $G_{2}=0.761$. When $G_{1}=G_{2}=G$, As Fig.3 shows, the maximum value changes to 1.03339 so long as $G=0.73$.

\begin{figure}[htbp]
      \centering
      \includegraphics[width=0.45\textwidth]{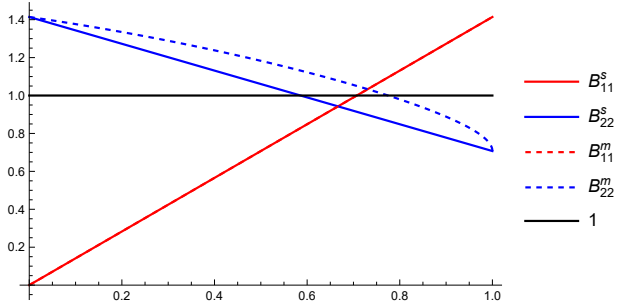}
      \caption{\small{Plot of BRGP quantities $B_{11}^s$ (red line), $B_{22}^s$ (blue line) $B_{11}^m$ (red dashed line), $B_{22}^m$ (blue dashed line) as functions of the precision factor G for $\mathrm{Alice_{1}-Bob-Charlie_{1}}$ and $\mathrm{Alice_{2}-Bob-Charlie_{2}}$ under the condition of $G_1$ = $G_2$ =$G$. $B_{11}^s$ and $B_{11}^m$ coincide because of the same pointer distribution. it's shown that $B_{11}^s$ and $B_{22}^s$ can't violate the BRGP inequality simultaneously with both $\mathrm{Alice_1}$ and $\mathrm{Charlie_{1}}$ implementing the square pointer type of weak measurements. $B_{11}^m$ and $B_{22}^m$ both exceed the bound of 1 in a narrow range when one chooses the square pointer and the other chooses the optimal pointer of weak measurements.}\label{B11B22}}
\end{figure}

\section{active network nonlocal sharing in the extended bilocal scenario}\label{active}

Different from passive network nonlocal sharing, when $\mathrm{Alice_1}$ and $\mathrm{Charlie_1}$ are willing to help $\mathrm{Alice_2}$ and $\mathrm{Charlie_2}$ exhibit network nonlocality as much as possible under the condition of guaranteeing the violation of BRGP inequality bewteen themselves, the nonlocal sharing emerging from this case is defined as active nonlocality sharing.

Assumed that all other conditions of the scenario remain unchanged, we can analyze active network nonlocality sharing by observing multiple violation of BRGP inequalities. Without loss of generality, we discuss the double violation of BRGP inequalities between $\mathrm{Alice_1}$-Bob-$\mathrm{Charlie_{1}}$ and $\mathrm{Alice_2}$-Bob-$\mathrm{Charlie_{2}}$. Firstly, we confirmed that active network nonlocality sharing can have a large range of double violation than that of passive nonlocal sharing numerically. But it is too complex to obtain a simple and distinct analytical result. When $G_{1}=G_{2}=G$, we could give a suboptimal analytical solution, which is a piecewise function. When G $\in [0,0.8]$, the maximal BRGP values of $\mathrm{Alice_1}$-Bob-$\mathrm{Charlie_{1}}$ and $\mathrm{Alice_2}$-Bob-$\mathrm{Charlie_{2}}$ are $B_{11}=\sqrt{2}G$ and $B_{22}=\frac{1+F}{\sqrt{2}}$ respectively with $\theta_{1,0}=\theta_{1,1}=\frac{\pi}{4}$ for $\mathrm{Alice_1}$ and $\mathrm{Charlie_1}$, and $\theta_{2,0}=\theta_{2,1}=\frac{\pi}{4}$ for $\mathrm{Alice_2}$ and $\mathrm{Charlie_2}$.
When G $\in (0.8,1]$, the suboptimal values of BRGP quantities $B_{11}, B_{22}$,
\begin{eqnarray}\label{BA1}
&&B_{11}=\frac{(F+\sqrt{2-F(2+F)})G}{\sqrt{2-2F}}\nonumber\\
&&B_{22}=\sqrt{1+F^3+\frac{1}{2}F^4} \geq 1.
\end{eqnarray}
when $\theta_{1,0}=\theta_{1,1}=\frac{1}{2}\arccos(1-\frac{F^2}{1-F})$ and $\theta_{2,0}=\theta_{2,1}=\arccos(\frac{2-F^2}{\sqrt{4+2F^3(2+F)}})$.

As illustrated in Fig.4, when the pointer distribution of the weak measurements for $\mathrm{Alice_1}$ and $\mathrm{Charlie_1}$ is optimal, we can find double violation of BRGP inequalities between $\mathrm{Alice}_1-\mathrm{Bob}-\mathrm{Charlie}_1$ and $\mathrm{Alice}_2-\mathrm{Bob}-\mathrm{Charlie}_2$ can exceed 1 simultaneously in the range of $G\in (\frac{1}{\sqrt{2}},1)$. Even as the precision factor $G$ approaches 1, network nonlocal sharing between $\mathrm{Alice}_1-\mathrm{Bob}-\mathrm{Charlie}_1$ and $\mathrm{Alice}_2-\mathrm{Bob}-\mathrm{Charlie}_2$ still can be observed.

\begin{figure}[htbp]
      \centering
      \includegraphics[width=0.45\textwidth]{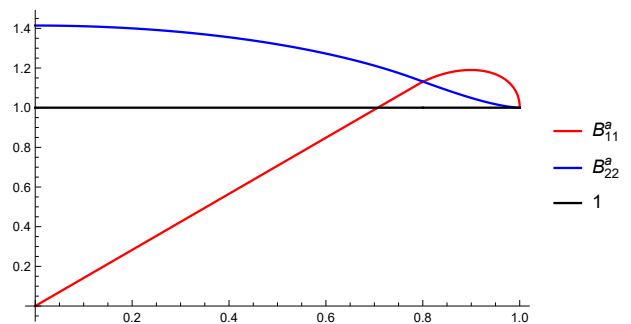}
      \caption{\small{Plot of BRGP quantities $B_{11}^a$ with the red line and $B_{22}^a$ with the blue line as functions of the precision factor G of $\mathrm{Alice}_1$ and $\mathrm{Charlie}_1$ for active network nonlocal sharing in an unbiased case. It can be seen that a double violation could happen even G is very close to 1.}\label{P1P2}}
 \end{figure}

\section{Noise resistance of network nonlocal sharing}\label{noise}
Network correlation in this scenario emerges when the central party which possesses two qubits
from two different sources performs a Bell-state measurement on them and nonlocality is generated between the
other two uncorrelated subsystems. Obviously, imperfect particle production reduces quantum network correlation. It is also interesting to discuss noise resistance of network nonlocal sharing in this case. Supposed that the shared states from these sources are not maximally entangled, but Werner states $\rho({v_{i}})$ with noise parameter $v_i$ of the form
\begin{eqnarray}\label{Bell state89}
\rho({v_{i}})=v_{i}(|\psi\rangle\langle \psi |) + \frac{1-v_{i}}{4}\mathbb{I}\otimes\mathbb{I}.
\end{eqnarray}
$v_{1}$ and $v_{2}$ are noise parameters for $\rho_{AB}$ and $\rho_{BC}$.

Assumed that all other conditions of the scenario remain unchanged, it is easy to obtain the BRGP quantities of $\mathrm{Alice}_1-\mathrm{Bob}-\mathrm{Charlie}_1$ and $\mathrm{Alice}_2-\mathrm{Bob}-\mathrm{Charlie}_2$
\begin{eqnarray}\label{B1}
&&B_{11}=\sqrt{2G_{1}G_{2}v_{1}v_{2}}\nonumber\\
&&B_{22}=\sqrt{\frac{(1+F_{1})(1+F_{2})v_{1}v_{2}}{2}}\nonumber
\end{eqnarray}
when $\theta_{1,0}=\theta_{1,1}=\frac{\pi}{4}$ for $\mathrm{Alice_1}$ and $\mathrm{Charlie_1}$, and $\theta_{2,0}=\theta_{2,1}=\frac{\pi}{4}$ for $\mathrm{Alice_2}$ and $\mathrm{Charlie_2}$.
Obviously, $V=\sqrt{v_{1}v_{2}}$ is a critical visibility for network nonlocality sharing. When $V\leq 0.884$, it is impossible to observe a double violation of BRGP inequalities between $\mathrm{Alice}_1-\mathrm{Bob}-\mathrm{Charlie}_1$ and $\mathrm{Alice}_2-\mathrm{Bob}-\mathrm{Charlie}_2$.

\section{Conclusion}

In this paper, network nonlocal sharing in the extended bilocal scenario has been completely discussed. It is clearly shown that network nonlocality can be revealed from the measurement results of any $\mathrm{Alice}_n-\mathrm{Bob}-\mathrm{Charlie}_m$ under appropriate measurement conditions, as all four BRGP inequalities based on $\mathrm{Alice}_n-\mathrm{Bob}-\mathrm{Charlie}_m$ can be violated simultaneously. It has a substantial divergence from Bell-type nonlocal sharing owing to that either the nonlocality sharing between $\mathrm{Alice_1}$-$\mathrm{Bob_{1}}$ and $\mathrm{Alice_2}$-$\mathrm{Bob_{2}}$ or the nonlocality sharing between $\mathrm{Alice_1}$-$\mathrm{Bob_{2}}$ and $\mathrm{Alice_2}$-$\mathrm{Bob_{1}}$ can never be observed in the Bell nonlocal sharing scenario \cite{Zhu,Cheng}. In particular, network nonlocal sharing can also be divided into two different types of nonlocality sharing, passive and active network nonlocal sharing, according to the different motivations of the former observers $\mathrm{Alice_1}$ and $\mathrm{Charlie_1}$. For active network nonlocal sharing, a double violation of BRGP inequalities can be always observed in a broad range which is extremely helpful for experimental realization. Besides, we have also studied the effect on network nonlocal sharing by using different pointer types of weak measurement. Finally, we analyzed noise resistance of network nonlocal sharing in the extended bilocal scenario.

Based on the current experimental techniques, the sharing of network nonlocality is experimentally observable. For instance, the recent implemented experimental schemes of entanglement swapping in optical systems \cite{Xu,Li} can be easily extended to observe the above conclusions. Weak measurement implementations therein can refer to recent experimental demonstration for Bell-type nonlocal sharing \cite{Hu,Schiavon,Feng,Zhu}. Our theoretical results provide a novel insight for understanding of network nonlocality.

\section{Acknowledgment}

This work was supported by the National Natural Science Foundation of China (Grant No. 12075245),the Natural Science Foundation of Hunan Province (2021JJ10033), the National Key R\&D Program of China ( 2017YFA0305000), the Fundamental Research Funds for the Central Universities, Xiaoxiang Scholars Programme of Hunan Normal university.


\end{document}